\documentclass[preprint]{aastex}
\usepackage{multirow}
\usepackage{graphicx}
\usepackage{color}

\shortauthors{Wang}

\begin{document}
\title{Accurate Group Delay Measurement for Radial Velocity Instruments Using the Dispersed Fixed Delay Interferometer Method}
\author{Ji Wang, Jian Ge, Xiaoke Wan, Brian Lee \& Nathan De Lee}
\affil{Department of Astronomy, University of Florida, Gainesville,
    FL, 32611}
\email{jwang@astro.ufl.edu}

\begin{abstract}

The Dispersed Fixed Delay Interferometer (DFDI) method is attractive for
its low cost, compact size and multi-object capability 
in precision radial velocity (RV) measurements. The phase shift of fringes of stellar
absorption lines is measured and then converted to an RV shift via an
important parameter, phase-to-velocity scale (PV scale) determined by
the group delay (GD) of a fixed delay interferometer. Two methods of GD measurement using a DFDI Doppler instrument
are presented in this paper: 1), GD measurement using white light
combs generated by the fixed delay interferometer; 2), GD calibration using an
RV reference star. These two methods provides adequate precision of GD measurement and calibration given current RV precision achieved by a DFDI Doppler instrument. They can potentially be used to measure GD of an fixed delay interferometer for sub-meter precision Doppler measurement with a DFDI instrument. Advantages and limitations of each method are discussed in details. The
two methods can serve as standard procedures of PV scale
calibration for DFDI instruments and cross checks for each other.

\end{abstract}


\section{Introduction}

As of Apr 2012, there are over 700 discovered exoplanets, and most of
them are detected by the radial velocity (RV)
technique\footnotemark\footnotetext{http://exoplanet.eu/;
http://exoplanets.org/}. RV precision of 1 $\rm{m}\cdot\rm{s}^{-1}$ has been routinely achieved~\citep{Bouchy2009,Howard2010} with instruments such as HARPS~\citep{Mayor2003} and HIRES~\citep{Vogt1994}, which are
 cross-dispersed echelle spectrographs. While cross-dispersed 
echelle spectrographs are commonly used in instruments for precision RV measurements, 
a method using a dispersed fixed delay interferometer 
(DFDI) has offered an alternative method~\citep{Ge2006,Fleming2010,Lee2011}. 
In this method, a Michelson-type interferometer is used in combination with 
a moderate resolution spectrograph, RV signals are then extracted from phase shift of interference
fringes of stellar absorption lines~\citep{Erskine2000,Ge2002b,Ge2002,Erskine2003}. 
 The details about the DFDI theory
and applications are discussed in~\citet{vanEyken2010} and~\citet{Wang2011}. 
Instrument adopting the DFDI method has demonstrated advantages such as low cost,
compact size and multi-object
capability~\citep{Ge2002,Ge2006,Fleming2010,Lee2011,Wisniewski2012}.

In the DFDI method, a fixed delay interferometer ~\citep{Wan2009b,Wan2011} plays a crucial role in creating stellar spectral 
fringes  for high precision RV
measurements~\citep{Ge2002,Erskine2003}. The Doppler sensitivity
can be optimized by carefully choosing the group delay (GD) of
the interferometer~\citep{Wang2011}. More specifically, GD of an interferometer should be chosen such that the spatial frequency of white light combs (WLCs) matches with that of a stellar spectrum after rotational broadening. GD is defined by the following equation:
\begin{equation}
\label{eq:GD_def} 
\rm{GD}(\nu)=-\frac{1}{2\pi}\cdot\frac{d\phi}{d\nu},
\end{equation}
where $\phi$ is phase shift and $\nu$ is optical frequency. The interferometer in a DFDI instrument is usually designed to be field-compensated to minimize the influence of input beam instability~\citep{Wan2009b,Wang2010}. It is realized by carefully selecting glass materials and thicknesses of two second surface mirrors such that their virtual images are overlapped. Because glasses are used in the optical paths,  $\phi$ does no longer linearly change with frequency, therefore GD is dependent on optical frequency. An inaccurate GD measurement may significantly limit
the RV measurement accuracy~\citep{Barker1974,vanEyken2010}. 

In practice, there may be several methods of measuring GD:
\begin{enumerate}
\item Calculate GD based on glass refractive indices using Sellmeier equation and thicknesses from manufacturer specification.
\item Forward model the spectrum of a known spectral source, such as an Iodine cell or a Th-Ar lamp.
\item Measure phase and frequency using a while light source, such as a tungsten lamp. 
\item Calibrate GD using a source with known velocity.
\end{enumerate}

Method 1 is straightforward but may lack of adequate precision because of uncertainty in parameters in Sellmeier equation and manufacturer tolerance for glass thickness. Method 2 holds great promise for accurately determining GD but there some current practical issues preventing us from adopting this method (see more detailed discussion in \S \ref{sec:WLCdisc}). We will use Method 3 and 4 to measure GD of an interferometer in this paper. 

\begin{figure}
\begin{center}
\includegraphics[width=12cm,height=9cm,angle=0]{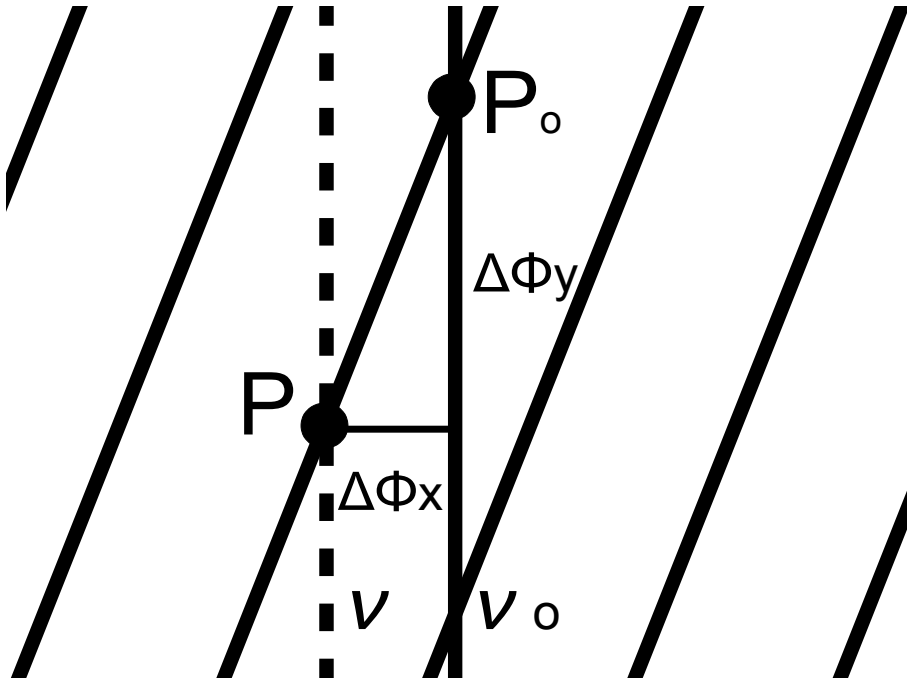}
\caption{Illustration of the DFDI method. Tilted lines represent interference combs generated by an interferometer. Vertical line represents an stellar absorption line (solid: original position with a frequency of $\nu_0$; dashed: shifted position with a frequency of $\nu$). \label{fig:DFDI_setup}}
\end{center}
\end{figure}

In the DFDI method,  GD determines the phase-to-velocity (PV) scale, the proportionality between the measured phase shift and the velocity
shift. Since the DFDI method is realized by coupling a fixed delay
interferometer with a post-disperser, the
resulting fringing spectrum$-$stellar absorption lines superimposing on the WLCs$-$is recorded on a CCD detector (illustrated in Fig. \ref{fig:DFDI_setup}). 
The fringe phase is expressed by the following equation:
\begin{equation}
\label{eq:fring_phase} \phi(\nu,y)=\frac{2\pi\cdot\tau(\nu,y)\cdot\nu}{c},
\end{equation}
where $y$ is the coordinate along slit direction, which is transverse to dispersion direction, $\tau$ is the optical path difference (OPD) of an interferometer and $c$ is the speed of light. Two mirrors (arms) of the interferometer are designed to be
tilted towards each other along the slit direction such that several fringes are formed
along each $\nu$ channel. The intersection of a stellar absorption line and a WLC moves (from $P_o$ to $P$ in Fig. \ref{fig:DFDI_setup}) if there is a shift of an absorption line due to a change of stellar RV. Consequently, a small change of $\phi$ in the dispersion direction, $\Delta\phi_x$, is induced:
\begin{eqnarray}
\label{eq:fring_phase_shift} \Delta\phi_x & = & \frac{d\phi}{d v}\cdot\Delta v=\frac{d\phi}{d\nu}\cdot\frac{d\nu}{d v}\cdot\Delta v \nonumber \\
& = & \frac{d\phi}{d\nu}\cdot\frac{\nu}{c}\cdot\Delta v=\Gamma\cdot\Delta v,
\end{eqnarray}
where $\Gamma$ is defined as phase-to-velocity scale (PV scale). It is determined by the GD of an interferometer, which becomes explicit if Equation (\ref{eq:GD_def}) and (\ref{eq:fring_phase_shift}) are combined:
\begin{equation}
\label{eq:PVS} \Gamma=-2\pi\cdot\rm{GD}\cdot\frac{\nu}{c}.
\end{equation}

At resolutions typically adopted by the DFDI method ($5,000\le R\le20,000$), stellar lines (line width$\sim$0.1$\AA$) are not resolved and a measurement of $\Delta\phi_x$ is extremely difficult. Instead, $\Delta\phi_y$, phase shift along $y$ direction can be measured, which is equal to $\Delta\phi_x$ if the combs generated by an interferometer are parallel to each other. This is a good approximation at very high orders of interference. The advantage of measuring $\Delta\phi_y$ instead of $\Delta\phi_x$ is seen from Fig. \ref{fig:DFDI_setup}, in which the physical shift in the $\nu$ direction is amplified in $y$ direction, the amplification rate is determined by the relative angle between the interferometer combs and a stellar absorption line. Therefore, $\Delta\phi_y$ is relatively easier to measure compared to $\phi_x$ and it is measured by fitting a well-sampled periodical flux signal along the $y$ direction in the DFDI method. Compared to conventional high-resolution Echelle method, the number of freedom for the DFDI method in the fitting process is much less and small Doppler phase shift can be relatively easier detected with a simple functional form, i.e., a sinusoidal function. However, we want to point out that while the DFDI method provides a boost in instrument Doppler sensitivity, the Doppler sensitivity is not strongly dependent on the amplification rate because flux slope decreases as amplification rate increases, which negates the gain of phase slope. 

The paper is organized as follows: in \S \ref{sec:OpdMeasurement}, we present
 a new method of GD measurement using a DFDI Doppler instrument, which exploits WLCs generated by a fixed delay interferometer. 
In \S \ref{sec:OPD_solution}, we present another method of GD calibration using an RV reference star. Observation results after implementing the newly measured GD are presented in \S \ref{sec:Implementation}. In \S \ref{sec:Conclusion}, we summarize and discuss the new results in this paper.

\section{GD Measurement Using White Light Combs}
\label{sec:OpdMeasurement}

\subsection{Method} \label{sec:OPDmethod}

MARVELS (Multi-object Apache Point Observatory Radial
Velocity Exoplanet Large-area Survey) is part of the Sloan Digital Sky Survey (SDSS) III~\citep{Gunn2006,Eisenstein2011}. The instrument covers a wavelength
range from 500 nm to 570 nm and uses a post-dispersive grating
with a spectral resolution of 11,000 after a fixed delay interferometer~\citep{Ge2009}. A Th-Ar emission
lamp and an iodine absorption cell serve as wavelength calibration
sources. The instrument setup of MARVELS~\citep{Ge2009,Wan2009b} is similar to the equipments that
measure GD as described in~\citet{Kovacs1995} and~\citet{Amotchkina2009}, in which a white light interferometer (WLI) is
combined with a post-disperser. However, the OPD is scanned by a moving picomotor in~\citet{Amotchkina2009} while it is realized by two relatively tilted arms in the WLC method using MARVELS instrument. WLCs are generated by the interferometer when fed by a white light source (e.g., a tungsten lamp). $\phi(\nu)$, the phase of each
frequency channel $\nu$, is measured and then unwrapped to remove
ambiguity of $2\pi$. GD is then derived by taking the derivative of
$\phi(\nu)$ according to Equation (\ref{eq:GD_def}).

\subsection{Data Reduction} \label{sec:DataReduction}

\begin{figure}
\begin{center}
\includegraphics[width=16cm,height=12cm]{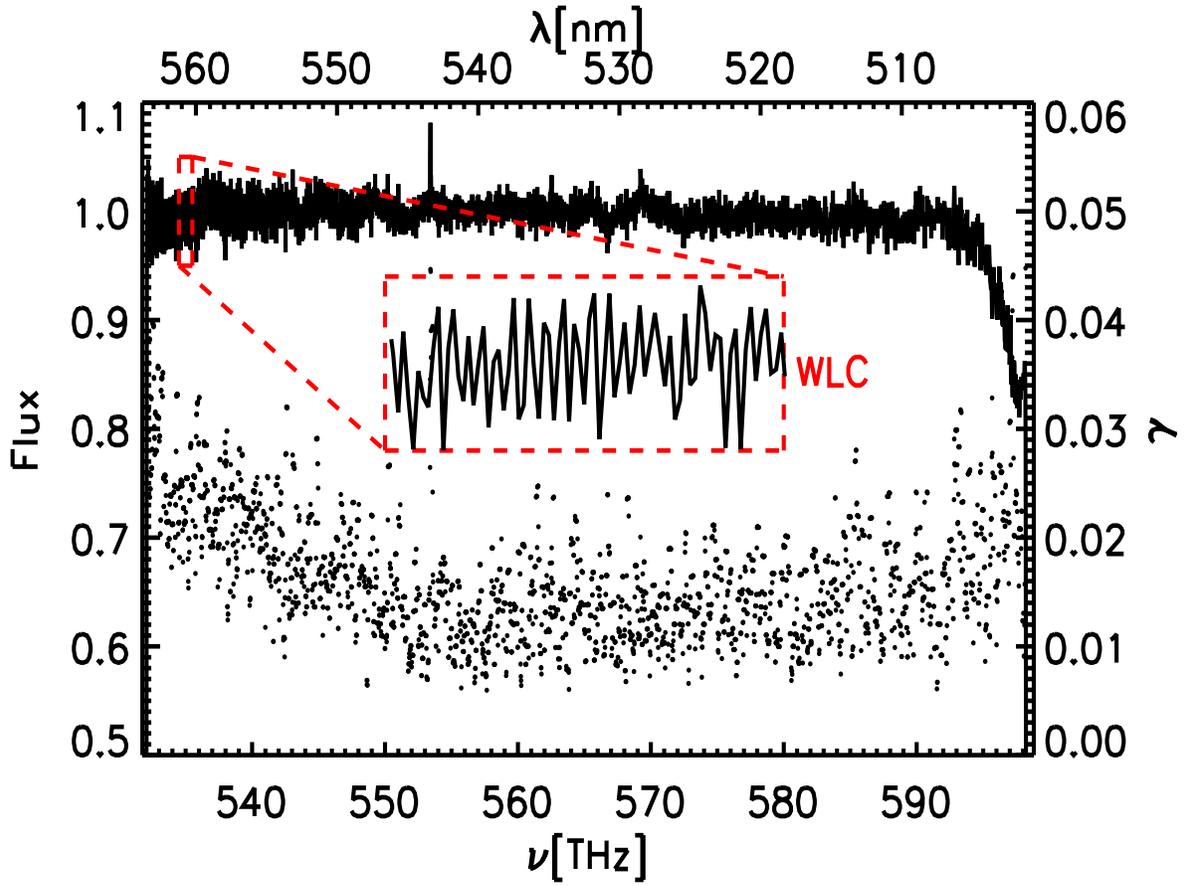} \caption{The normalized flux and visibility ($\gamma$) as a function of frequency of 
a tungsten spectrum taken with MARVELS. The solid line is the normalized flux and filled circles represent visibilities in different frequency channels. \label{fig:Flux_Vis_Fre}}
\end{center}
\end{figure}

Standard spectroscopy reduction procedures are performed with an IDL
data reduction pipeline dedicated to MARVELS. Fig \ref{fig:Flux_Vis_Fre} shows an example of normalized flux as a function of frequency for a processed spectrum. A 
zoom-in sub-plot shows the WLCs produced by frequency modulation of the interferometer. Visibility, defined as the ratio of half of peak-valley value to the DC offset, increases with frequency in the red part of the spectrum. The increasing visibility in the blue end of the spectrum is not physical but caused by  an increasing photon noise and our algorithm of visibility calculation.

The fringe phases as a function of $\nu$ are calculated by the Hilbert
transform technique~\citep{Rochford1999}: the signal
$H(\nu)$ is obtained by firstly removing the negative Fourier
components of $F(\nu)-$the flux distribution with frequency$-$and then conducting an inverse Fourier
transform. The phases $\phi(\nu)$ are obtained by calculating and unwrapping the arguments of $H(\nu)$. We find that the phase change between pixels exceeds $\pi$ in the blue part and therefore the phase unwrap cannot be successfully applied, so we decide to use only part of the spectrum with a pixel range from 1800 to 3800 for phase unwrapping. A third-order polynomial is used to fit $\phi$ as a function $\nu$. GD is then calculated according to its definition (Equation (\ref{eq:GD_def})).

\subsection{GD Measurement Results} \label{sec:MarvelsResult}

\begin{figure}
\begin{center}
\includegraphics[width=9cm,height=12cm,angle=0]{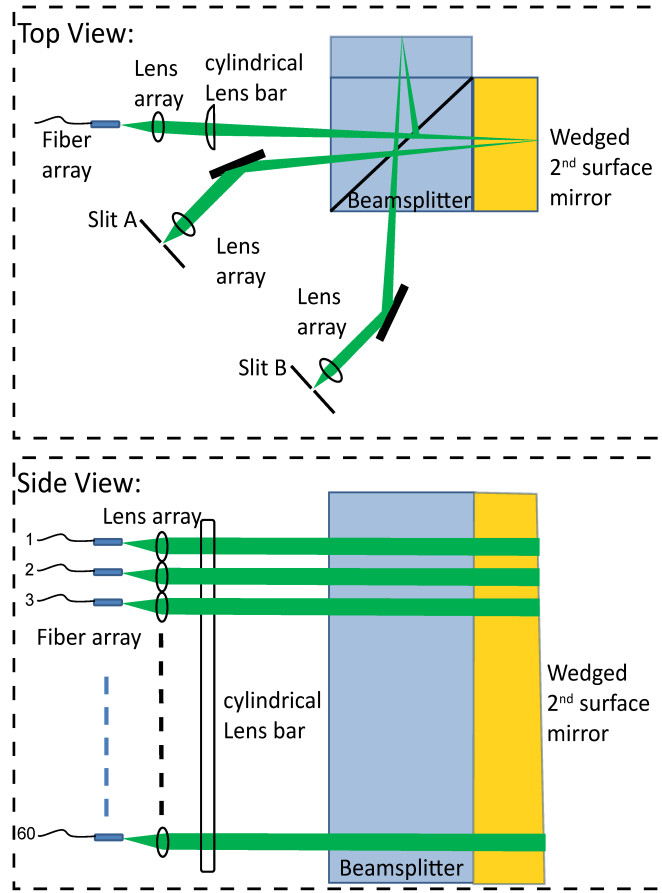}
\caption{Top: top view of an individual fiber beam feeding of the MARVELS interferometer. Two spectra are formed by one fiber. One (Slit A) is from the returning beam arm while the other one (Slit B) is from the forwarding beam arm. Bottom: side view of the fiber array beam feeding of the MARVELS interferometer. There are 60 fibers yielding 120 spectra. Note the exaggerated wedge angle of the shown second surface mirror, GD gradually changes along the vertical direction.\label{fig:illus_TopView}}
\end{center}
\end{figure}

\begin{figure}
\begin{center}
\includegraphics[width=16cm,height=12cm]{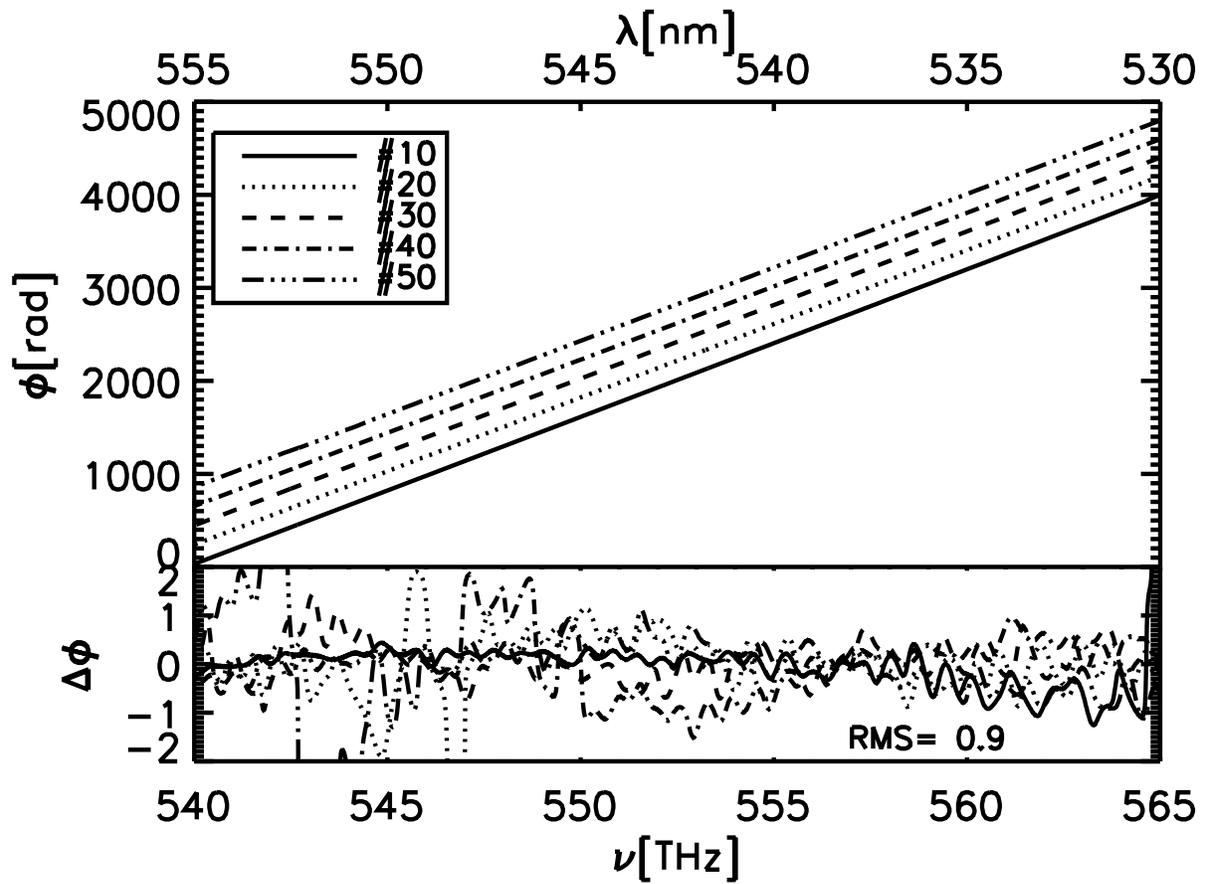} \caption{Top: white light combs phase as a function of frequency at different fiber locations. Bottom: phase residual after third-order polynomial fitting. \label{fig:PH_Fre_FiberNum}}
\end{center}
\end{figure}

\begin{figure}
\begin{center}
\includegraphics[width=16cm,height=12cm]{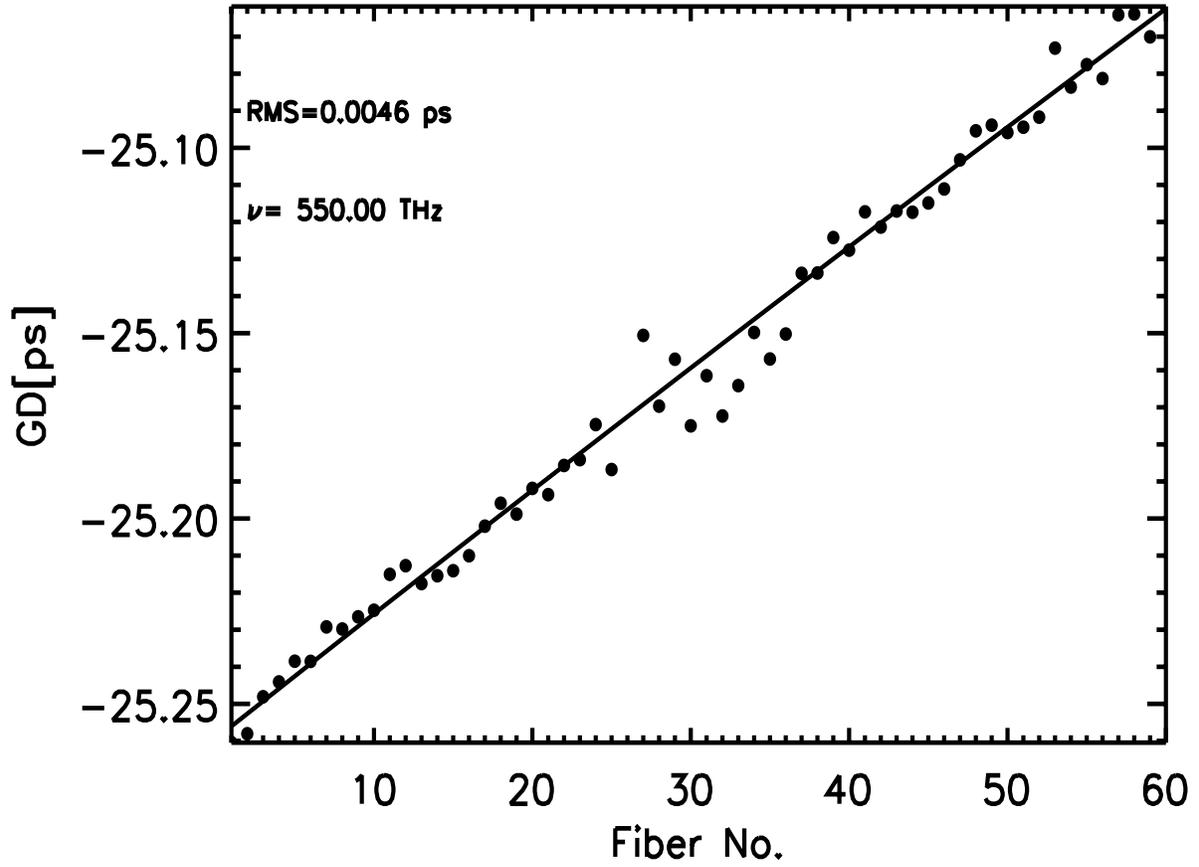} \caption{Measured group delay as a function fiber number. Filled circles are measured results, solid line represents the best second-order polynomial fitting with an RMS fitting error of 0.0046 ps. Measurement results can be found in Table \ref{tab:GD_err_fre} at other frequencies. \label{fig:GD_SpecNum}}
\end{center}
\end{figure}

\begin{figure}
\begin{center}
\includegraphics[width=16cm,height=12cm]{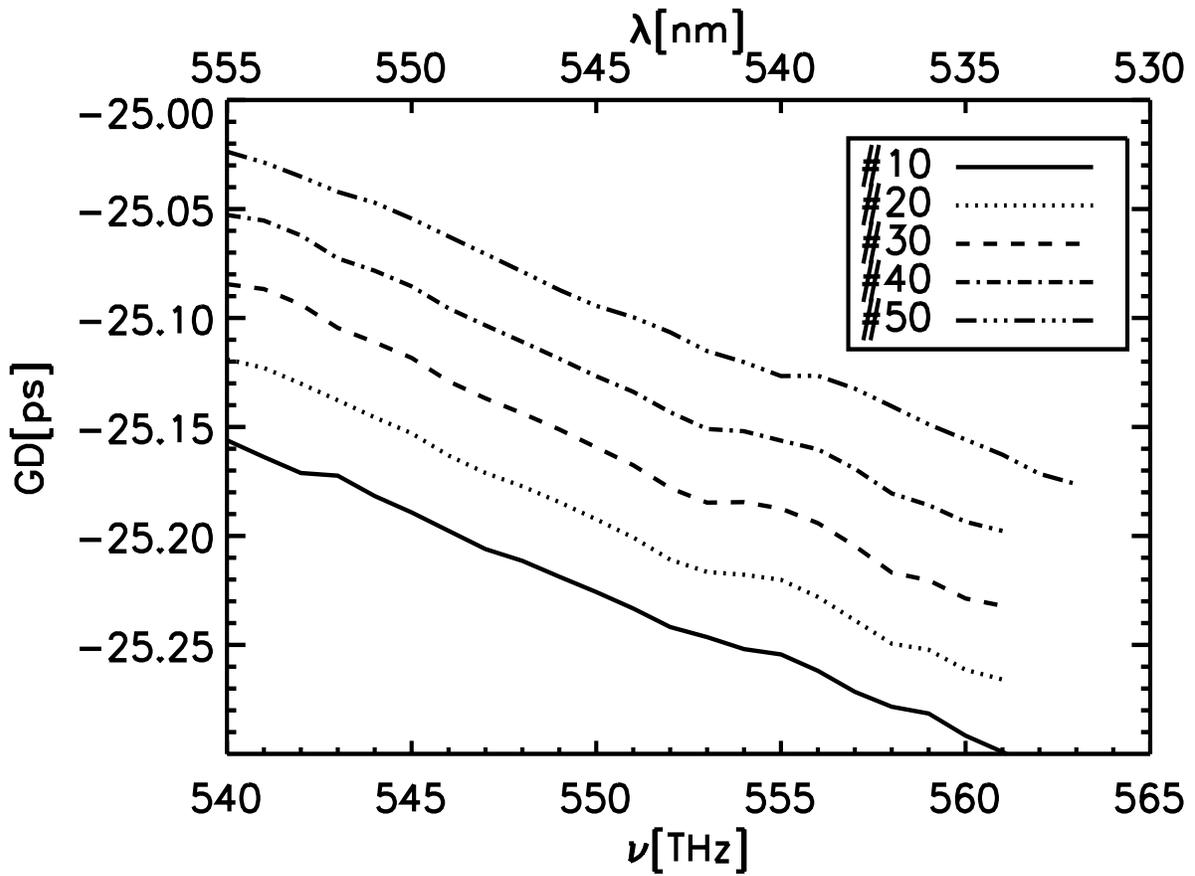} \caption{GD as a function of frequency at different fiber numbers. GD measurement results vs. frequency and fiber numbers can be found in Table \ref{tab:GD_err_fre}. \label{fig:GD_Fre_FiberNum}}
\end{center}
\end{figure}

The top view and side view of the MARVELS interferometer are shown in Fig. \ref{fig:illus_TopView}.  60 fibers are mounted and each creates two spectra, one is picked from the forwarding beam and the other one is from the returning beam (see Fig. \ref{fig:illus_TopView}). In total, 120 spectra are formed, allowing us to measure GD at 60 positions on the interferometer along vertical (slit) direction. Each position corresponds to a fiber number. There are 24 pixels along the slit direction for each spectrum. We chose 15 rows in the middle to measure GD because of relatively higher photon flux, and thus smaller photon noise in the middle region of the spectrum. The top panel of Fig. \ref{fig:PH_Fre_FiberNum} shows phase measurement results for center row as a function of frequency at different fiber numbers. Phase fitting residual (shown on the bottom panel of Fig. \ref{fig:PH_Fre_FiberNum}, RMS=0.9 rad) is consistent with photon-noise limited measurement error (see \S \ref{sec:Err_Ana} for details). GD for a particular fiber number is obtained by averaging the results of GD measurements for those rows associated with the fiber. Fig. \ref{fig:GD_SpecNum} shows the results at $\nu$=550 THz as a function of fiber number. Note that the two arms of the interferometer are intentionally tilted to each other and the 60 fibers are evenly mounted along the slit direction. The measured GDs should gradually vary with fiber number. We use a second-order polynomial to fit the GD variation with the fiber number. The fitting residual has an RMS of 0.0046 ps. Fig. \ref{fig:GD_Fre_FiberNum} shows fitted GD as a function of frequency for different fibers. GD varies 0.15 ps (0.6\%) across measurement range from 540 to 565 THz. Ignoring GD dependence of frequency would result in 180 $\rm{m}\cdot\rm{s}^{-1}$ measurement offset between two ends of measurement range (assuming a true RV of 30,000 $\rm{m}\cdot\rm{s}^{-1}$, which is a typical stellar RV value due to the Earth's barycentric motion). Table \ref{tab:GD_err_fre} provides the polynomial fitting coefficients of GD vs. fiber number at different frequencies within measurement range. 

\subsection{GD Measurement Error Analysis} \label{sec:Err_Ana}

Two physical parameters, $\phi$ and $\nu$, are measured in the experiment. The
uncertainty of the $\phi$ measurement is $\sim$0.8 rad under photon-noise limited condition assuming a S/N of 120 and a typical fringe visibility of 1.5\%. The uncertainty due to the wavelength calibration
is $\sim$0.002 THz (0.02\AA). We conduct a bootstrapping process to investigate the
uncertainty of GD caused by the measurement uncertainties of
$\phi$ and $\nu$. We add gaussian noises with standard deviation of
measurement errors to both $\phi$ and $\nu$ and calculate the group
delay. We run 1000 iterations for bootstrapping in order to estimate the
uncertainty of GD. The median of the relative error of GD measurements, \ $\delta \rm{GD}/\rm{GD}$, is $4.4\times10^{-5}$. In
comparison, the median of the relative GD measurement error is $\sim1.8\times10^{-4}$ after smoothing by fitting a polynomial to GD variation with the fiber number. This number does not agree with the relative GD error predicted by the bootstrapping experiment. We suspect that the
uncounted error in the bootstrapping simulation comes from image distortion due to optics which the data pipeline has not fully corrected for, e.g., spectrum curvature, spectral line slant, etc. In a 2-D
spectrum as illustrated in Fig. \ref{fig:DFDI_setup}, the phase shift between adjacent pixels along slit direction is $\sim$0.6 rad, and the
phase shift between each wavelength chanel is $\sim$2.5 rad. An imperfect spectrum
curvature tracing tends to shift pixel in the slit direction while
an imperfect slant correction can affect pixel shifting in both slit and
dispersion directions. The range of unwrapped phase is $\sim$4000 rad. For one fiber, if a gradually-changing phase error is introduced by the data pipeline when correcting for the optical distortion, for example, 0.4 rad deviation from true value at one end while no deviation at the other end, a relative error of GD would be caused with an estimation of  $0.4/4000=1\times10^{-4}$. If different fibers are treated independently, which is the case for the MARVELS data reduction pipeline, then this gradually-changing phase error, introduced by imperfect optical distortion correction, may explain the standard deviation error we see after the polynomial fitting for GDs as a function of fiber number.

\section{GD Calibration: Observing an RV Reference Star} \label{sec:OPD_solution}

\subsection{Method} \label{sec:method_ref}

A deviated PV scale would result in an inaccurate velocity measurement given the same amount of fringe phase shift:
\begin{equation}
\label{eq:Delta_phi_GD}
\Delta\phi=\Delta v^\prime\cdot\Gamma^\prime=\Delta v\cdot\Gamma,
\end{equation}
where $\Delta v^\prime$ represents a measured velocity shift while $\Delta v$ represents a true velocity shift. Combining Equation (\ref{eq:PVS}) and (\ref{eq:Delta_phi_GD}), we obtain the following equation from which GD can be calculated by using the measured velocity shift of an object with a known velocity.
\begin{equation}
\label{eq:GD_cal_ref}
\rm{GD}=\frac{\rm{GD}^\prime\cdot\Delta v^\prime}{\Delta v}.
\end{equation}
This approach is similar to that of~\citet{Barker1974}, but the difference is that the latter applied correction for discrete laser frequencies while we seek corrections for a continuous frequency distribution. 

In order to realize the method of GD calibration using an RV reference star, we need to: 1), assume a GD$^\prime$ that is close to the true value of GD; 2), measure velocity shift $\Delta v^\prime$ based on an assumed GD$^\prime$; 3), know the true value of the velocity shift of an RV reference star.

\subsection{GD Calibration Precision} \label{sec:precision_ref}

The calibration precision using an RV reference star is determined by measurement error of $\Delta v^\prime$:
\begin{equation}
\label{eq:GD_cal_err_ref}
\delta \rm{GD}=\frac{\rm{GD}^\prime\cdot\delta v^\prime}{\Delta v},
\end{equation}
where $\delta v^\prime$ is RV measurement uncertainty.~\citet{Wang2011} provided a method of calculating photon-limited RV uncertainty for the DFDI method. Under photon-noise limited condition, we expect the GD calibration error to be determined by the photon-limited RV uncertainty within the instrument band width, which is provided in Table \ref{tab:TeffDeltaV_S_offset}. In Equation (\ref{eq:GD_cal_err_ref}), GD$^\prime$ is usually estimated to be within a few percent of true GD, $\Delta v$ is statistically $\sim$10,000 $\rm{m}\cdot\rm{s}^{-1}$ given a uniform reference star distribution and a quarter year observational availability. Therefore, relative error of GD measurement is $\sim2\times10^{-4}$ for an RV reference star with a $T_{\rm{eff}}$ of 4500 K. 

\section{Implementation of Measured GD in Astronomical Observations} \label{sec:Implementation}

We use an RV reference star, HIP 14810 (V=8.5), as an example to show the RV measurement results after implementation of the newly measured GD using the WLC method. HIP 14810 is a star known to harbor 3 planets and its RV jitter is estimated to be 2 $\rm{m}\cdot\rm{s}^{-1}$~\citep{Wright2009}. After RV changes due to instrument drift, the Earth's barycentric motion and orbiting planets are removed, RV RMS error is 17.13 $\rm{m}\cdot\rm{s}^{-1}$ but has not reached the predicted photon-limited RV uncertainty (4.8 $\rm{m}\cdot\rm{s}^{-1}$, S/N=80 with a half wavelength coverage from 535 to 565 nm). RV RMS error is expected to be further reduced after the data pipeline is improved in the future.

\begin{figure}
\begin{center}
\includegraphics[width=16cm,height=12cm]{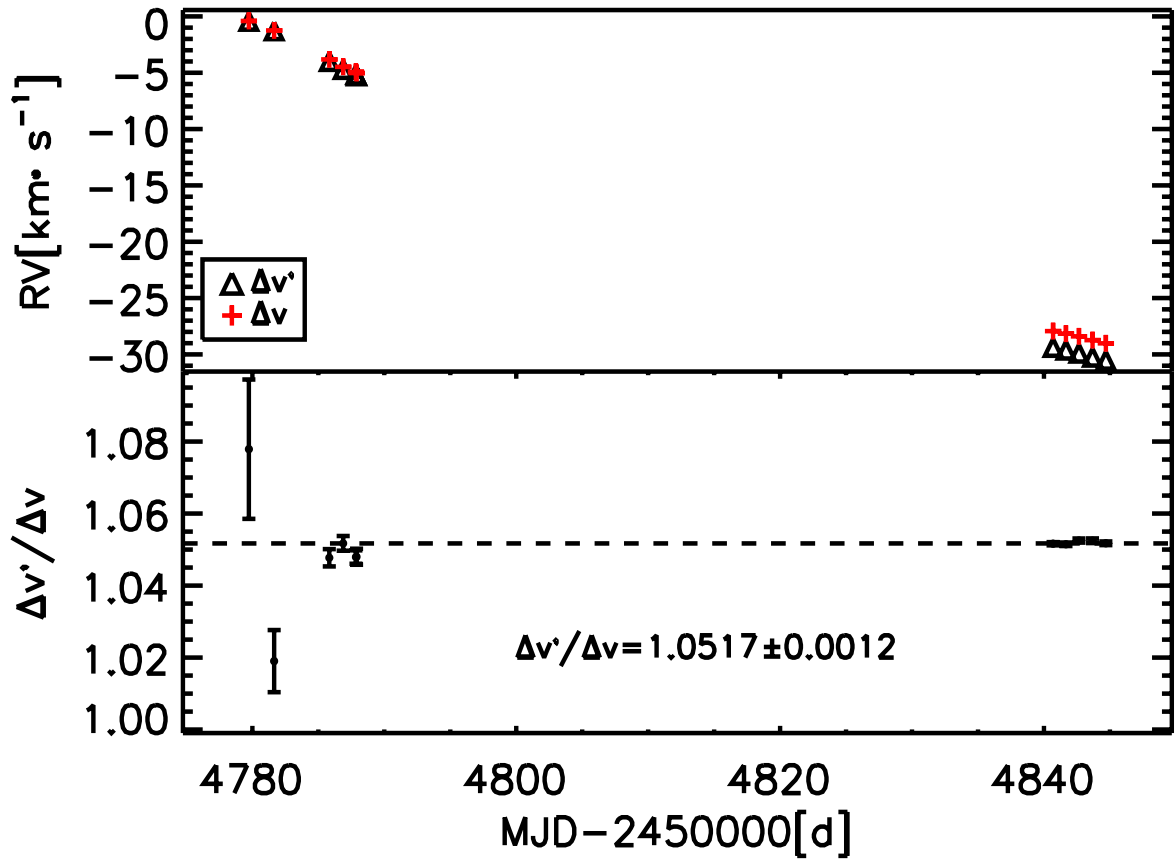} \caption{Top: measured ($\Delta v^\prime$) and true ($\Delta v$) RVs of HIP 14810 (barycentric velocity not corrected) over a period of 70 days. Bottom: the ratio of $\Delta v^\prime$ and $\Delta v$ as a function of time. \label{fig:HIP_14810_before}}
\end{center}
\end{figure}

We also examine the reference star GD calibration method. We use one spectral block within measurement range centering at 550 THz (540-560 THz) and set GD$^\prime$ to be an arbitrary value of -23.873 ps. The measured RVs (barycentric velocity not corrected) are shown in Fig. \ref{fig:HIP_14810_before}. After applying correction according to Equation (\ref{eq:GD_cal_ref}), we find that the GD is -25.107$\pm$0.027 ps. In comparison, GD measurement result of fiber number 51 (the fiber for HIP 14810) using the WLC method gives -25.091$\pm$0.005 ps (refer to Table \ref{tab:GD_err_fre}). We confirm that the GDs measured by these two methods are consistent with each other at 68\% significance level.

\section{Summaries and Discussions}
\label{sec:Conclusion}

\subsection{Summaries}

The PV scale is an important parameter in the DFDI method that translates a
measured phase shift to an RV shift, and is determined by the group
delay (GD) of an interferometer. We have provided and discussed two
methods of GD measurement and calibration: 1), GD measurement
using white light combs (WLCs) generated by the interferometer in a DFDI Doppler
instrument; 2), GD calibration using an RV reference star (RS). Table \ref{tab:Cal_Comp} summarizes the main results and the comparison between these two methods. The accuracy of GD measurement is sufficient for current RV precision achieved with instruments using the DFDI method~\citep{Fleming2010, Lee2011, Muirhead2011}. However, higher measurement and calibration precision is required in the near future as higher RV precision is achieved by DFDI instruments in search for exoplanets. RS and WLC methods can serve as complementary methods of GD measurement and calibration for DFDI instruments. 

\subsection{Discussions}
\subsubsection{White Light Comb (WLC) Method}
\label{sec:WLCdisc}

The GD measurement using WLCs created by the
interferometer provides a direct way of calibrating the PV scale. In
the region where combs are visible, effective S/N is relatively low ($\sim$15) because of  low comb visibility (1.5\%). In addition, GD cannot be measured in the region where combs are not visible, which limits the application of
this method. We are able to measure GD in a region that accounts for half of the spectrum coverage. Extrapolation beyond the measurement range may result in
large uncertainties. The major issue facing the method is that the
data reduction pipeline may have introduced unknown errors while correcting
optical distortions such as spectrum curvature and slant. 

In principle, we can use a tungsten lamp with an iodine cell or a Th-Ar lamp instead of a tungsten lamp in order to increase the fringe visibility. However, there are some practical concerns that hinder us from applying the above solutions: 1), line blending, because of low spectral resolution, many spectral lines can not be resolved and it is not certain at this stage how line blending affects phase measurement; 2), illumination correction, which is required to correct for illumination profile in slit direction in order to properly measure the phase. We adopt a self-illumination correction procedure in the pipeline which requires a certain continuum level to be successfully achieved. Th-Ar is less affected by line blending if a careful line selection process is involved, but it does not have enough continuum level for self-illumination correction. An experiment is being conducted in which a second interferometer is used to improve the visibility of WLCs so that GD is more precisely measured at a higher effective S/N for a wider frequency coverage.

When compared to previous work in the field of GD measurement,~\citet{Amotchkina2009} achieved a measurement precision of $1\times10^{-4}$ ps. Our measurement of GD has a typical accuracy of $\sim6\times10^{-3}$ ps (limited by effective S/N and systematic errors), which is more than an order of magnitude lower. However, it is shown in \S \ref{sec:Err_Ana} that substantial improvement would be able to be achieved once the data reduction pipeline has a better handle of optical distortion.~\citet{Wan2010} measured GD for the MARVELS interferometer using a scanning WLI method and achieved a precision of $0.6\times10^{-5}\sim2.4\times10^{-5}$ ps, which is more than two order of magnitude better than the results in this paper. However, there are practical concerns using GD measurement results from the scanning WLI method because they are not measured in situ, therefore it is not easy to associate a position in a WLI measurement to a fiber position.

\subsubsection{Reference Star (RS) Method}

The GD calibration using an RV reference star (RS) is a
self-calibrating process and has the potential of achieving a high calibration precision if the following requirements are met: 1), the RV reference star has a large velocity shift during a observation window; 2), the RV reference star is bright; 3), the data reduction
pipeline is able to produce the photon-limited RV precision. In addition, GD is practically measured within a certain band width:
\begin{equation}
\label{eq:GD_cal_ref_wb}
\rm{GD}(\nu)=\frac{\int_{\Delta\nu}\rm{GD}(\nu)\omega(\nu)d\nu}{\int_{\Delta\nu}\omega(\nu)d\nu},
\end{equation}
where $\omega(\nu)$ is weight function. The band width, $\Delta\nu$, should be small such that the dispersion effect is negligible. The limitations stated above prevent us from precisely determining the PV scale at the position of each fiber because not every fiber has a continuous observation on a bright known RV reference star. For MARVELS, the brightest RV reference star available has
V mag of 8 and the resulting S/N is $\sim$100 per pixel. However, the RS method is a very promising approach for a single-object DFDI instrument because only one bright reference star
is required in the field. We are planning to apply this method in
calibrating GD of another DFDI instrument (EXPERT) at KPNO
2.1m telescope~\citep{Ge2010}. Note that the S/N can be further increased by conducting multiple independent measurements and increasing instrument throughput.

We would like to express our deepest gratitude to the anonymous referee of this paper, without whom there would not have been this paper. We  acknowledge the support from NSF with grant NSF AST-0705139, NASA with
grant NNX07AP14G (Origins), UCF-UF SRI program, DoD ARO Cooperative Agreement W911NF-09-2-0017, SDSS III consortium, Dharma Endowment Foundation and the University of Florida. 

Funding for SDSS-III (\url{http://www.sdss3.org/}) has been provided by the Alfred P. Sloan Foundation, the Participating Institutions, the National Science Foundation, and the U.S. Department of Energy Office of Science.

SDSS-III is managed by the Astrophysical Research Consortium for the Participating Institutions of the SDSS-III Collaboration including the University of Arizona, the Brazilian Participation Group, Brookhaven National Laboratory, University of Cambridge, Carnegie Mellon University, University of Florida, the French Participation Group, the German Participation Group, Harvard University, the Instituto de Astrofisica de Canarias, the Michigan State/Notre Dame/JINA Participation Group, Johns Hopkins University, Lawrence Berkeley National Laboratory, Max Planck Institute for Astrophysics, Max Planck Institute for Extraterrestrial Physics, New Mexico State University, New York University, Ohio State University, Pennsylvania State University, University of Portsmouth, Princeton University, the Spanish Participation Group, University of Tokyo, University of Utah, Vanderbilt University, University of Virginia, University of Washington, and Yale University.

\bibliographystyle{apj}
\bibliography{mybib_JW_XW}{}

\begin{thebibliography}{28}
\expandafter\ifx\csname natexlab\endcsname\relax\def\natexlab#1{#1}\fi

\bibitem[{Amotchkina {et~al.}(2009)Amotchkina, Tikhonravov, Trubetskov, Grupe,
  Apolonski, \& Pervak}]{Amotchkina2009}
Amotchkina, T.~V., Tikhonravov, A.~V., Trubetskov, M.~K., Grupe, D., Apolonski,
  A., \& Pervak, V. 2009, Appl. Opt., 48, 949

\bibitem[{{Barker} \& {Schuler}(1974)}]{Barker1974}
{Barker}, L.~M., \& {Schuler}, K.~W. 1974, Journal of Applied Physics, 45, 3692

\bibitem[{{Bouchy} {et~al.}(2009){Bouchy}, {Mayor}, {Lovis}, {Udry}, {Benz},
  {Bertaux}, {Delfosse}, {Mordasini}, {Pepe}, {Queloz}, \&
  {Segransan}}]{Bouchy2009}
{Bouchy}, F., {et~al.} 2009, A\&A, 496, 527

\bibitem[{{Eisenstein} {et~al.}(2011){Eisenstein}, {Weinberg}, {Agol},
  {Aihara}, {Allende Prieto}, {Anderson}, {Arns}, {Aubourg}, {Bailey},
  {Balbinot}, \& et~al.}]{Eisenstein2011}
{Eisenstein}, D.~J., {et~al.} 2011, \aj, 142, 72

\bibitem[{{Erskine}(2003)}]{Erskine2003}
{Erskine}, D.~J. 2003, PASP, 115, 255

\bibitem[{{Erskine} \& {Ge}(2000)}]{Erskine2000}
{Erskine}, D.~J., \& {Ge}, J. 2000, in Astronomical Society of the Pacific
  Conference Series, Vol. 195, Imaging the Universe in Three Dimensions, ed.
  {W.~van Breugel \& J.~Bland-Hawthorn}, 501--+

\bibitem[{{Fleming} {et~al.}(2010){Fleming}, {Ge}, {Mahadevan}, {Lee},
  {Eastman}, {Siverd}, {Gaudi}, {Niedzielski}, {Sivarani}, {Stassun},
  {Wolszczan}, {Barnes}, {Gary}, {Cuong Nguyen}, {Morehead}, {Wan}, {Zhao},
  {Liu}, {Guo}, {Kane}, {van Eyken}, {De Lee}, {Crepp}, {Shelden}, {Laws},
  {Wisniewski}, {Schneider}, {Pepper}, {Snedden}, {Pan}, {Bizyaev},
  {Brewington}, {Malanushenko}, {Malanushenko}, {Oravetz}, {Simmons}, \&
  {Watters}}]{Fleming2010}
{Fleming}, S.~W., {et~al.} 2010, \apj, 718, 1186

\bibitem[{{Ge}(2002)}]{Ge2002}
{Ge}, J. 2002, ApJ, 571, 165

\bibitem[{{Ge} {et~al.}(2002){Ge}, {Erskine}, \& {Rushford}}]{Ge2002b}
{Ge}, J., {Erskine}, D.~J., \& {Rushford}, M. 2002, PASP, 114, 1016

\bibitem[{{Ge} {et~al.}(2006){Ge}, {van Eyken}, {Mahadevan}, {DeWitt}, {Kane},
  {Cohen}, {Vanden Heuvel}, {Fleming}, {Guo}, {Henry}, {Schneider}, {Ramsey},
  {Wittenmyer}, {Endl}, {Cochran}, {Ford}, {Mart{\'{\i}}n}, {Israelian},
  {Valenti}, \& {Montes}}]{Ge2006}
{Ge}, J., {et~al.} 2006, \apj, 648, 683

\bibitem[{{Ge} {et~al.}(2009){Ge}, {Lee}, {de Lee}, {Wan}, {Groot}, {Zhao},
  {Varosi}, {Hanna}, {Mahadevan}, {Hearty}, {Chang}, {Liu}, {van Eyken},
  {Wang}, {Pais}, {Chen}, {Shelden}, \& {Costello}}]{Ge2009}
{Ge}, J., {et~al.} 2009, in Presented at the Society of Photo-Optical
  Instrumentation Engineers (SPIE) Conference, Vol. 7440, Society of
  Photo-Optical Instrumentation Engineers (SPIE) Conference Series

\bibitem[{{Ge} {et~al.}(2010){Ge}, {Zhao}, {Groot}, {Chang}, {Varosi}, {Wan},
  {Powell}, {Jiang}, {Hanna}, {Wang}, {Pais}, {Liu}, {Dou}, {Schofield},
  {McDowell}, {Costello}, {Delgado-Navarro}, {Fleming}, {Lee}, {Bollampally},
  {Bosman}, {Jakeman}, {Fletcher}, \& {Marquez}}]{Ge2010}
{Ge}, J., {et~al.} 2010, in Presented at the Society of Photo-Optical
  Instrumentation Engineers (SPIE) Conference, Vol. 7735, Society of
  Photo-Optical Instrumentation Engineers (SPIE) Conference Series

\bibitem[{{Gunn} {et~al.}(2006){Gunn}, {Siegmund}, {Mannery}, {Owen}, {Hull},
  {Leger}, {Carey}, {Knapp}, {York}, {Boroski}, {Kent}, {Lupton}, {Rockosi},
  {Evans}, {Waddell}, {Anderson}, {Annis}, {Barentine}, {Bartoszek}, {Bastian},
  {Bracker}, {Brewington}, {Briegel}, {Brinkmann}, {Brown}, {Carr},
  {Czarapata}, {Drennan}, {Dombeck}, {Federwitz}, {Gillespie}, {Gonzales},
  {Hansen}, {Harvanek}, {Hayes}, {Jordan}, {Kinney}, {Klaene}, {Kleinman},
  {Kron}, {Kresinski}, {Lee}, {Limmongkol}, {Lindenmeyer}, {Long}, {Loomis},
  {McGehee}, {Mantsch}, {Neilsen}, {Neswold}, {Newman}, {Nitta}, {Peoples},
  {Pier}, {Prieto}, {Prosapio}, {Rivetta}, {Schneider}, {Snedden}, \&
  {Wang}}]{Gunn2006}
{Gunn}, J.~E., {et~al.} 2006, \aj, 131, 2332

\bibitem[{{Howard} {et~al.}(2010){Howard}, {Marcy}, {Johnson}, {Fischer},
  {Wright}, {Isaacson}, {Valenti}, {Anderson}, {Lin}, \& {Ida}}]{Howard2010}
{Howard}, A.~W., {et~al.} 2010, Science, 330, 653

\bibitem[{Kov\'{a}cs {et~al.}(1995)Kov\'{a}cs, Osvay, Bor, \&
  Szip\"{o}cs}]{Kovacs1995}
Kov\'{a}cs, A.~P., Osvay, K., Bor, Z., \& Szip\"{o}cs, R. 1995, Opt. Lett., 20,
  788

\bibitem[{{Lee} {et~al.}(2011){Lee}, {Ge}, {Fleming}, {Stassun}, {Gaudi},
  {Barnes}, {Mahadevan}, {Eastman}, {Wright}, {Siverd}, {Gary}, {Ghezzi},
  {Laws}, {Wisniewski}, {Porto de Mello}, {Ogando}, {Maia}, {Nicolaci da
  Costa}, {Sivarani}, {Pepper}, {Cuong Nguyen}, {Hebb}, {De Lee}, {Wang},
  {Wan}, {Zhao}, {Chang}, {Groot}, {Varosi}, {Hearty}, {Hanna}, {van Eyken},
  {Kane}, {Agol}, {Bizyaev}, {Bochanski}, {Brewington}, {Chen}, {Costello},
  {Dou}, {Eisenstein}, {Fletcher}, {Ford}, {Guo}, {Holtzman}, {Jiang}, {French
  Leger}, {Liu}, {Long}, {Malanushenko}, {Malanushenko}, {Malik}, {Oravetz},
  {Pan}, {Rohan}, {Schneider}, {Shelden}, {Snedden}, {Simmons}, {Weaver},
  {Weinberg}, \& {Xie}}]{Lee2011}
{Lee}, B.~L., {et~al.} 2011, \apj, 728, 32

\bibitem[{{Mayor} {et~al.}(2003){Mayor}, {Pepe}, {Queloz}, {Bouchy},
  {Rupprecht}, {Lo Curto}, {Avila}, {Benz}, {Bertaux}, {Bonfils}, {Dall},
  {Dekker}, {Delabre}, {Eckert}, {Fleury}, {Gilliotte}, {Gojak}, {Guzman},
  {Kohler}, {Lizon}, {Longinotti}, {Lovis}, {Megevand}, {Pasquini}, {Reyes},
  {Sivan}, {Sosnowska}, {Soto}, {Udry}, {van Kesteren}, {Weber}, \&
  {Weilenmann}}]{Mayor2003}
{Mayor}, M., {et~al.} 2003, The Messenger, 114, 20

\bibitem[{{Muirhead} {et~al.}(2011){Muirhead}, {Edelstein}, {Erskine},
  {Wright}, {Muterspaugh}, {Covey}, {Wishnow}, {Hamren}, {Andelson}, {Kimber},
  {Mercer}, {Halverson}, {Vanderburg}, {Mondo}, {Czeszumska}, \&
  {Lloyd}}]{Muirhead2011}
{Muirhead}, P.~S., {et~al.} 2011, PASP, 123, 709

\bibitem[{Rochford \& Dyer(1999)}]{Rochford1999}
Rochford, K.~B., \& Dyer, S.~D. 1999, J. Lightwave Technol., 17, 831

\bibitem[{{van Eyken} {et~al.}(2010){van Eyken}, {Ge}, \&
  {Mahadevan}}]{vanEyken2010}
{van Eyken}, J.~C., {Ge}, J., \& {Mahadevan}, S. 2010, \apjs, 189, 156

\bibitem[{{Vogt} {et~al.}(1994){Vogt}, {Allen}, {Bigelow}, {Bresee}, {Brown},
  {Cantrall}, {Conrad}, {Couture}, {Delaney}, {Epps}, {Hilyard}, {Hilyard},
  {Horn}, {Jern}, {Kanto}, {Keane}, {Kibrick}, {Lewis}, {Osborne},
  {Pardeilhan}, {Pfister}, {Ricketts}, {Robinson}, {Stover}, {Tucker}, {Ward},
  \& {Wei}}]{Vogt1994}
{Vogt}, S.~S., {et~al.} 1994, in Society of Photo-Optical Instrumentation
  Engineers (SPIE) Conference Series, Vol. 2198, Society of Photo-Optical
  Instrumentation Engineers (SPIE) Conference Series, 362--+

\bibitem[{Wan {et~al.}(2011)Wan, Ge, \& Chen}]{Wan2011}
Wan, X., Ge, J., \& Chen, Z. 2011, Appl. Opt., 50, 4105

\bibitem[{Wan {et~al.}(2009)Wan, Ge, Wang, \& Lee}]{Wan2009b}
Wan, X., Ge, J., Wang, J., \& Lee, B. 2009in  (SPIE), 742406

\bibitem[{Wan {et~al.}(2010)Wan, Wang, \& Ge}]{Wan2010}
Wan, X., Wang, J., \& Ge, J. 2010, Appl. Opt., 49, 5645

\bibitem[{{Wang} {et~al.}(2011){Wang}, {Ge}, {Jiang}, \& {Zhao}}]{Wang2011}
{Wang}, J., {Ge}, J., {Jiang}, P., \& {Zhao}, B. 2011, ApJ, 738, 132

\bibitem[{{Wang} {et~al.}(2010){Wang}, {Wan}, \& {Ge}}]{Wang2010}
{Wang}, J., {Wan}, X., \& {Ge}, J.~C. 2010, in Society of Photo-Optical
  Instrumentation Engineers (SPIE) Conference Series, Vol. 7734, Society of
  Photo-Optical Instrumentation Engineers (SPIE) Conference Series

\bibitem[{{Wisniewski} {et~al.}(2012){Wisniewski}, {Ge}, {Crepp}, {De Lee},
  {Eastman}, {Esposito}, {Fleming}, {Gaudi}, {Ghezzi}, {Gonzalez Hernandez},
  {Lee}, {Stassun}, {Agol}, {Allende Prieto}, {Barnes}, {Bizyaev}, {Cargile},
  {Chang}, {Da Costa}, {Porto De Mello}, {Femen{\'{\i}}a}, {Ferreira}, {Gary},
  {Hebb}, {Holtzman}, {Liu}, {Ma}, {Mack}, {Mahadevan}, {Maia}, {Nguyen},
  {Ogando}, {Oravetz}, {Paegert}, {Pan}, {Pepper}, {Rebolo}, {Santiago},
  {Schneider}, {Shelden}, {Simmons}, {Tofflemire}, {Wan}, {Wang}, \&
  {Zhao}}]{Wisniewski2012}
{Wisniewski}, J.~P., {et~al.} 2012, \aj, 143, 107

\bibitem[{{Wright} {et~al.}(2009){Wright}, {Fischer}, {Ford}, {Veras}, {Wang},
  {Henry}, {Marcy}, {Howard}, \& {Johnson}}]{Wright2009}
{Wright}, J.~T., {et~al.} 2009, \apjl, 699, L97

\end{thebibliography}


\clearpage

%

\clearpage

\begin{deluxetable}{ccccc}

\tablewidth{400pt}

\tablecaption{GD measurement results as a function of spectrum number ($GD(\#)=C_0+C_1\cdot\#+C_2\cdot\#^2$) and standard deviation ($\delta$GD) at different frequencies ($\nu$)}

\tablehead{\colhead{$\nu$[THz]} & \colhead{$C_0$} & \colhead{$C_1$}  & \colhead{$C_2$} & \colhead{$\delta GD$[ps]}  }

\startdata 
\label{tab:GD_err_fre}
540.0000 & -2.5195975233e+01 & 4.1224184409e-03 & -1.3537701123e-05 & 0.0066 \\
542.0000 & -2.5216854974e+01 & 4.8064043246e-03 & -2.3438315106e-05 & 0.0070 \\
544.0000 & -2.5219400268e+01 & 3.8524896642e-03 & -8.1355543995e-06 & 0.0057 \\
546.0000 & -2.5232735338e+01 & 3.5346857088e-03 & -2.6247780042e-06 & 0.0055 \\
548.0000 & -2.5246276599e+01 & 3.5171101494e-03 & -3.3394045215e-06 & 0.0047 \\
550.0000 & -2.5259435353e+01 & 3.3877963986e-03 & -1.7477210448e-06 & 0.0046 \\
552.0000 & -2.5270775315e+01 & 2.8029235840e-03 & 9.6338400722e-06 & 0.0066 \\
554.0000 & -2.5286851225e+01 & 3.5335302125e-03 & -4.0412480867e-06 & 0.0093 \\
556.0000 & -2.5295872014e+01 & 3.4000221425e-03 & -2.5813725398e-07 & 0.0067 \\
558.0000 & -2.5303654859e+01 & 2.3432563708e-03 & 1.8389109388e-05 & 0.0087 \\
560.0000 & -2.5319347252e+01 & 2.6464319719e-03 & 1.2493738694e-05 & 0.0075 \\

\enddata

\end{deluxetable}


\begin{table}
\caption{MARVELS predicted RV uncertainty (at an average S/N of 100) vs. $T_{\rm{eff}}$\label{tab:TeffDeltaV_S_offset}}
\begin{tabular}{ccccccc}
\hline
$T_{\rm{eff}}$[$K$] & 4500 & 5000 & 5500 & 6000 & 6500 & 7000 \\
\hline
$\delta v^\prime$[$\rm{m}\cdot\rm{s}^{-1}$] & 1.9 & 2.3 & 2.7 & 3.1 & 3.5 & 4.0 \\
\hline
\end{tabular}
\end{table}

\begin{table}
\caption{Comparison between two methods of GD measurement and calibration
\label{tab:Cal_Comp}}
\begin{tabular}{ccc}
\hline
 & WLC & RS \\
\hline

Spectrum Coverage & Half & Full  \\
S/N & $\sim$15 & 100 for V$\sim$8$^a$  \\
Current precision & $4.6\times10^{-3}\sim9.3\times10^{-3}$ ps & 0.027 ps \\
Current RV error$^b$ &   $\sim$ 2 $\rm{m}\cdot\rm{s}^{-1}$ & 10.8 $\rm{m}\cdot\rm{s}^{-1}$ \\
Potential precision & $\sim3.1\times10^{-4}$ ps & $\sim5\times10^{-3}$ ps \\
Potential RV error$^b$ & $\sim$ 0.1 $\rm{m}\cdot\rm{s}^{-1}$ & $\sim$ 2 $\rm{m}\cdot\rm{s}^{-1}$ \\
Dependence on observation & $\times$ & $\checkmark$ \\
Dependence on pipeline & $\checkmark$ & $\checkmark$ \\

\hline

\hline
\end{tabular}

\tablecomments{a: assuming MARVELS throughput; b: RV error is calculated assuming a true velocity shift ($\Delta v$) of 10,000 $\rm{m}\cdot\rm{s}^{-1}$ according to Equation (\ref{eq:GD_cal_ref}).}

\end{table}

\end{document}